\documentclass[acmsmall]{acmart}
\AtBeginDocument{%
  \providecommand\BibTeX{{%
    \normalfont B\kern-0.5em{\scshape i\kern-0.25em b}\kern-0.8em\TeX}}}

\setcopyright{acmcopyright}
\copyrightyear{2018}
\acmYear{2018}
\acmDOI{XXXXXXX.XXXXXXX}

\begin{document}

\title{Towards a Unified Pandemic Management Architecture: Survey, Challenges and Future Directions}


\author[1]{Satyaki Roy}
\affiliation{%
  \institution{University of North Carolina, Chapel Hill}
  \country{USA}
  \authornote{Satyaki Roy and Nirnay Ghosh contributed equally.}}
\email{satyakir@unc.edu}

\author[1]{Nirnay Ghosh}
\affiliation{%
 \institution{Indian Institute of Engineering Science and Technology (IIEST), Shibpur}
 \country{India}}

\author[2]{Nitish Uplavikar}
\affiliation{%
  \institution{University of Missouri, Columbia}
  \country{USA}}

\author[3]{Preetam Ghosh}
\affiliation{%
 \institution{Virginia Commonwealth University, Virginia}
 \country{USA}}


\begin{abstract}

The pandemic caused by SARS-CoV-2 has left an unprecedented impact on health, economy and society worldwide. Emerging strains are making pandemic management increasingly challenging. There is an urge to collect epidemiological, clinical, and physiological data to make an informed decision on mitigation measures. Advances in the Internet of Things (IoT) and edge computing provide solutions for pandemic management through data collection and intelligent computation. While existing data-driven architectures attempt to automate decision-making, they do not capture the multifaceted interaction among computational models, communication infrastructure, and the generated data. In this paper, we perform a survey of the existing approaches for pandemic management, including online data repositories and contact-tracing applications. We then envision a unified pandemic management architecture that leverages the IoT and edge computing to automate recommendations on vaccine distribution, dynamic lockdown, mobility scheduling and pandemic prediction. We elucidate the flow of data among the layers of the architecture, namely, cloud, edge and end device layers. Moreover, we address the privacy implications, threats, regulations, and existing solutions that may be adapted to optimize the utility of health data with security guarantees. The paper ends with a lowdown on the limitations of the architecture and research directions to enhance its practicality.

\end{abstract}

\keywords{Pandemic Management, Internet of Things, Optimization, Machine Learning, Privacy}


\maketitle

\section{Introduction} \label{intro}
COVID-19 is an infectious disease that has had major health, economic and social impact on the world. It has infected over 293 million people and claimed nearly 5.5 million lives globally. The manifestation of COVID-19 ranges from asymptomatic/mild symptoms to severe respiratory illness, hospitalization and death, while the survivors are afflicted with several long-term effects~\cite{lopez2021more}. The efficacy of vaccines are challenged by the rapid and unabated emergence of new strains varying in virulence and transmissibility~\cite{NIH-Closer,shin2020covid}. COVID-19 is expected to acquire an endemic status over time, causing intermittent outbreaks. Hence, it is imperative to design policies around the clinical management strategies, rehabilitation measures, human behavior and pandemic response~\cite{phillips2021coronavirus,anderson2020developing,boseNLP,roy2021identifying}. 

The proliferation of digital technology has resulted in the generation of volumes of pandemic-related data made available in the form of public repositories. These clinical, epidemiological, physiological, socioeconomic, genomic, etc. data not only disseminate information on the latest developments, but have opened up a new vista of computational research on COVID-19~\cite{napolitano2021impact}. This research aims to create management plans to combat epidemic outbreaks by optimized allocation of human and material resources such as manufacturing production, human labor, transportation, etc.~\cite{queiroz2020impacts}. Furthermore, it identifies the limitations in the existing technological deployments and ways to make them more practicable~\cite{gupta2021future}.   
These computational frameworks propose application of a wide array of methodologies such as artificial intelligence, machine learning, deep learning, optimization, statistics, network science, epidemiology, bioinformatics, etc. to meet two primary goals~\cite{kautish2021computational}. First, they inform the researchers, policymakers and general public of the spatio-temporal and behavioral trends that are associated with the onset of severe outbreaks~\cite{hu2021human}. Second, they recommend the decisions that will contribute towards curbing contagion~\cite{raza2021introduction}. For example, these models may prescribe the distribution of vaccines among zones based on socioeconomic, demographic factors and human mobility to avoid infection spread or the effective duration and timing of lockdowns to cut down on mortality and hospitalization as well as economic losses~\cite{yakovyna2020recommendation}. 

\begin{figure}[h!]
	\centering
	\includegraphics[width=2.5in]{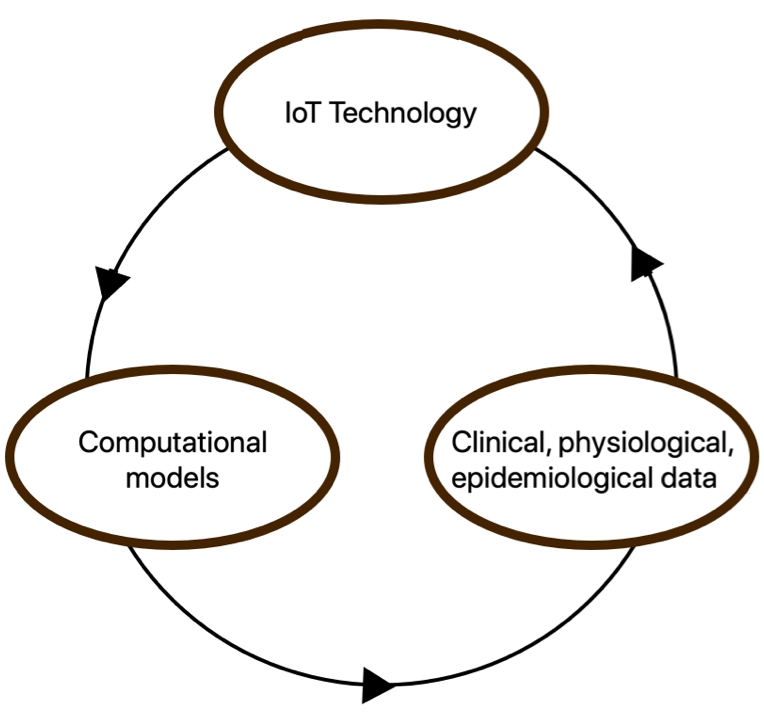}
	\caption{The proposed pandemic management architecture presents a holistic picture of the multilayered interaction among the epidemiological, physiological and clinical data, methods, and the IoT communication infrastructure.}
	\label{fig:cycle}
\end{figure}

The advent of the Internet of Things (IoT) has revolutionized the field of healthcare. We have seen a barrage of healthcare solutions that rely on the mobility, clinical, physiological, and epidemiological data generated by the IoT devices~\cite{yousif2021iot}. Pandemic management would undoubtedly have been more challenging in the absence of these innovations in digital technology. At the same time, the abundance of IoT devices and the data generated therefrom creates a problem of plenty, necessitating a comprehensive \textit{pandemic management architecture}. This architecture should provide a blueprint for the (1) communication technologies for the collection of data using citizens' personalized devices as well as data repositories, (2) data offloading from user equipment to compute/storage infrastructures, and (3) the use of computational models for data analysis to enable informed decision-making~\cite{jahmunah2021future}. 

Efforts have been made to design data-driven architectures that automate decision-making regarding pandemic healthcare. Shamout et al. introduced a deep learning architecture to learn pulmonary complications due to COVID-19~\cite{shamout2021artificial}. Karaarslan et al. proposed an epidemic management system that employs data to distribute resources effectively, while enabling the interaction of multiple platforms and meeting scalability and interoperability issues~\cite{karaarslan2021artificial}. Alam discussed a blockchain-based mobile healthcare architecture for real-time monitoring of the COVID-19 patients~\cite{alam2021blockchain}, while Otoom et al. explained an IoT framework that processes the real-time symptom data to run diagnostics and monitoring~\cite{otoom2020iot}. \textit{However, these frameworks do not provide a unified picture of the multilayered interaction among the data, methods, and the wireless communication infrastructure of a realistic pandemic management architecture} (see Fig. \ref{fig:cycle}).

\begin{table}[h!]
    \centering
\begin{tabular}{|c|c|}
    \hline
    \textbf{Term} & \textbf{Abbreviation}	\\
    \hline
    Attribute-based Access Control & ABAC \\
    \hline 
    Contagion Potential & CP	\\
    \hline
    Differential Privacy & DP	\\
    \hline
    Federated Learning & FL	\\
    \hline
    Graph Attention Network & GAT \\ 
    \hline
    Giant Connected Component & GCC	\\
    \hline
    General Data Protection Regulation & GDPR \\
    \hline 
    Grant Recurrent Network & GRU	\\
    \hline
    Internet of Things & IoT	\\
    \hline
    Internet Protocol & IP	\\
    \hline
    Internet Service Protocol & ISP	\\
    \hline
    Local Area Network & LAN \\ 
    \hline
    Long-Term Evolution & LTE \\ 
    \hline
    Mobile Edge Computing & MEC \\ 
    \hline 
    Machine Learning as a service & MLaaS \\ 
    \hline 
    Oblivious Transfer & OT \\ 
    \hline 
    Organization-based Access Control & OrBAC \\ 
    \hline 
    Private Information Retrieval & PIR \\ 
    \hline 
    Radio Access Network & RAN \\ 
    \hline 
    Role-based Access Control & RBAC \\ 
    \hline 
    Radio Network Controller & RNC \\
    \hline
    Received Signal Strength Indicator & RSSI \\ 
    \hline 
    Susceptible-Exposed-Infected-Recovered-Death & SEIRD \\ 
    \hline
    Single Instruction Multiple Data & SIMD \\
    \hline
    Spatiotemporal Attention Network & SPAN \\ 
    \hline
    User Equipment & UE \\
    \hline 
    \end{tabular}
    \caption{List of abbreviations}
    \label{tab:abb}
\end{table}

In this paper, we propose an architecture that consolidates the features of the existing computational frameworks into IoT and edge computing technologies. It will gather sensory physiological, clinical and epidemiological data from the general public, harness computational approaches to process ground and historical data, and send periodic recommendations to civic authorities and participating users. The \textit{pandemic management architecture} operates on three levels, namely, individuals possessing smart devices (which we refer to as \textit{IoT devices}) infused with data analytic capabilities, a cloud layer comprising servers enabled with high data storage and processing power, and an intermediate edge layer offering on-demand computation, communication, and caching services to applications running on IoT devices. Hence, the requests from IoT devices are not directly sent to the back-end cloud but are offloaded to intermediate edge nodes that are at the edge of the users' networks, resulting in improved latency and better bandwidth utilization for the rest of the Internet. 

We discuss the relevant studies on the online data repositories and mobile contact-tracing applications as well as four representative pandemic management tasks, namely, vaccine distribution, dynamic lockdown, contact tracing, and pandemic prediction (see Sec. \ref{sec:survey}). Next, we cover the three layers of the architecture, followed by how the edge computation layer pulls time-series data from cloud-hosted repositories as well as the mobile devices, runs them through the computational models, and recommends human actions regarding vaccines, social distancing, lockdowns, etc. (see Sec. \ref{sec:arc}). We explore the privacy concerns of sharing of confidential data and related strategies to achieve high data utility and necessary data anonymity (see Sec. \ref{sec:priv}). Finally, we highlight the shortcomings of the pandemic framework that hinder practical implementation and motivate several new research directions (see Sec. \ref{sec:disc}).

This paper is organized as follows. Sec. II deals with the preliminary concepts relevant to the subsequent discussion. Sec. III presents a survey on the computational frameworks on COVID-19. Sec. IV discusses the proposed pandemic management architecture. Finally, in Sec. V and VI, we cover the challenges, future directions and conclusions, respectively. (Refer to Table \ref{tab:abb} for the list of abbreviations.)

\section{Preliminaries}

\subsection{Vaccine Distribution}\label{sec:dist}
\begin{figure}[h!]
	\centering
	\includegraphics[width=2.5in]{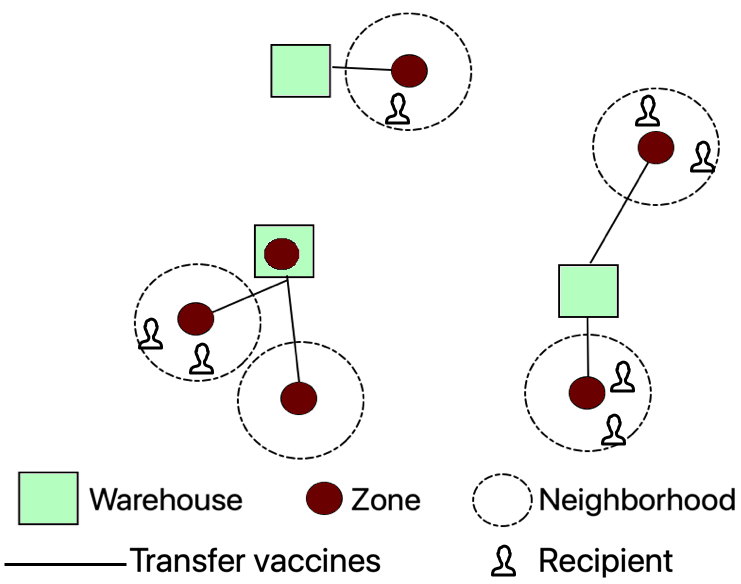}
	\caption{Vaccine warehouse, zones and neighborhood}
	\label{fig:vacc}
\end{figure}

\noindent The vaccines are housed in the warehouses and transferred to the vaccination sites of the zones (see Fig. \ref{fig:vacc}). (The warehouses can have their own sites as well.) The vaccine recipients from the neighborhood visit the sites to get inoculated~\cite{roy2021optimal,xu2021hub}.  

\subsection{Epidemic Model}\label{sec:seird}
\noindent The susceptible-exposed-infected-recovered-death (SEIRD) epidemic model is applied to the mathematical modelling of infectious diseases~\cite{hethcote2000mathematics}. The \textit{susceptible} ($S$) class comprises individuals who are not exposed to the infection. Once exposed to infected individuals, they may transfer to the \textit{exposed} ($E$) category, and this transition is controlled by a rate $\beta$. The $E$ class comprises asymptomatic or untested individuals, who transition to the (tested) \textit{infected} ($I$) class with probability $\sigma$. The individuals in $I$ transition to another state with a probability $\gamma$; this other state can be either \textit{recovered} ($R$) or \textit{dead} ($D$) with probabilities $1 - \alpha$ and $\alpha$, respectively, as shown below. Note that $\beta = \gamma \times R_0$, where $R_0$ is the basic reproduction number (defined as the number of new infections a single infectious individual creates in a susceptible population~\cite{linka2020reproduction}) that has a median value of $3$, but can be equal to $5.7$ or even more as per previous literature~\cite{korolev2020identification,sanche2020early}. Thus, unlike, ($\gamma, \rho, \alpha$), $\beta$ is not a transition probability.

\begin{align}
\centering
    & S \xrightarrow[I]{\beta} E \label{eq:seird1}\\
    & E \xrightarrow{\sigma} I \label{eq:seird2}\\
    & I \xrightarrow{\gamma \times (1 - \alpha)} R \label{eq:seird3}\\
    & I \xrightarrow{\gamma \times \alpha} \label{eq:seird4}D 
\end{align}

\subsection{Graph Theory}\label{sec:graph}

\noindent A graph is an ordered pair $G = (V, E)$ where $V$ is a finite, non-empty set of objects called \textit{vertices} (or nodes); and $E$ is a (possibly empty) set of 2-subsets of $V$, called \textit{edges} \cite{newman2003structure}. A directed graph is a graph in which edges have directions. A directed edge $(u,v) \in E$, allows unidirectional information flow from vertex $u$ to $v$ and not necessarily from $v$ to $u$. In a weighted graph, $(u, v) \in E$ is associated with a weight $w_{u,v} \in [0, 1]$, measuring the strength of influence of $u$ on $v$.

\subsection{Components of Wireless Communication Architecture}
\noindent The major components in our envisioned pandemic management architecture are as follows:
\begin{enumerate}
    \item \textit{\underline{Core network}}: It specifies the backbone network for the Internet, which interconnects geographically dispersed computer networks and constitutes several core routers. The core network provides data routes for exchange of information between different Local Area Networks (LANs) and subnetworks at geographically dispersed locations. The core routers are owned by different cellular companies and are managed by regional and national Internet Service Providers (ISPs). Each router has multiple high-speed interfaces for faster exchange of Internet Protocol (IP) packets among communicating networks and supports a wide range of routing protocols. 

    \item \textit{\underline{Cloud servers}}: Cloud computing is a computing paradigm that offers on-demand services to the end-users through a pool of computing resources that includes storage services, computing resources, and so on~\cite{mell2011nist}. It constitutes high-performance computing clusters hosted in a remote network and managed by infrastructure service providers. These servers handle application requests from the end device users, and to cater to tens of thousands of user requests simultaneously, they have significant capacities of computation, storage, and network bandwidths. Each cloud server can host multiple application containers/content servers that handle requests in parallel. When the data processing is complete, the cloud servers will send the results back to the end-users via the backhaul network(s). Moreover, using the storage services, these servers can store massive volumes of time-series data obtained at regular intervals from/by APIs managed by external data sources. 
    
    \item \textit{\underline{Radio Access Network (RAN)}}: A radio access network (RAN) is a part of the mobile telecommunication system, and it provides a predefined range of frequencies (frequency spectrum) for enabling wireless communication between mobile phones or any wireless controlled machines (laptops, tablets, wearables, etc.) with the mobile core network~\cite{park2012performance}. With the accustomed support of mobile Internet using RAN long-term evolution (LTE), 4G/5G now has an edge over other wireless mobile telecommunication technologies providing the best user experience~\cite{khan20094g}. RAN covers a wide geographical area divided into cells, and each cell is integrated with its base station. Base stations are typically connected via microwave or landlines. Each base station has a radio network controller (RNC) that carries out mobile management functions and connects the former with backhaul networks through the Ethernet interface, which supports a high data transfer rate~\cite{park2012performance}. Mobile Edge Computing (MEC) -- a combination of mobile devices, applications and the data they consume and produce -- offers cloud computing capabilities within the RAN by pushing cloud resources to the mobile edge. Thus, instead of allowing direct mobile traffic between the core network and the end-user, MEC connects the user to the nearest cloud service-enabled edge network. Deploying MEC at the base station enhances computation and avoids bottlenecks and system failure~\cite{luan2015fog,abbas2017mobile}. 
    
    \item \textit{\underline{Users' Equipment (UE)}}: Users’ equipment consists of mobile, handheld devices such as smartphones, laptops, tablet computers, and wearables. These devices are equipped with multiple sensors that can be invoked opportunistically by mobile applications for collecting environmental data. Some of the sensory data that can be obtained are of location, motion, gesture, audio, light, etc. These devices are intelligent as they can receive feedback from applications and perform subtle actuation to adapt to varying environmental conditions. Besides this, users can also participate in data acquisition tasks by voluntarily sharing their personal or environmental data. Such devices communicate with remote content and compute servers using Wi-Fi, LTE/4G/5G technologies, and can communicate with neighboring devices using short-range communication protocols such as BLE/Bluetooth, ZigBee, and Wi-Fi direct protocols.
    
    \item \textit{\underline{Mobile Edge Computing (MEC) servers}}: Mobile edge computing (MEC) constitutes geo-distributed servers with built-in IT services~\cite{abbas2017mobile}. These servers have fair amounts of computing, storage, and communication bandwidth. MEC servers may be deployed at a fixed location, such as at an LTE base station or a multi-technology (3G/LTE) cell aggregation site~\cite{luan2015fog}. MEC may utilize cellular network elements, such as the base station, Wi-Fi access point, or femto access point (i.e., low power cellular base station)~\cite{abbas2017mobile}. The fundamental value proposition of MEC is that it pushes cloud capabilities (compute, storage, communication) to the edge of the users’ network. It enables most applications (running on UEs) to offload general-purpose requests to the nearby MEC servers rather than sending them to the remote cloud. Such offloading helps to fulfill the QoS requirements of resource-intensive, delay-sensitive, and high-bandwidth demanding applications running on end devices. Additionally, the bandwidth bottleneck situations at backhaul network(s) between the end-users and cloud servers are also avoided. However, the MEC server forwards them to the cloud in case of very high compute-intensive or large data processing requests.
 \end{enumerate}

\section{Computational Pandemic Management Frameworks}\label{sec:survey}
\noindent In this section, we survey the literature on online repositories, contact-tracing mobile applications, as well as the current representative computational frameworks for COVID-19 management.

\subsection{Online Repositories}\label{sec:repo}
    \noindent We discuss standard databases that are periodically updated with emerging information on different facets of COVID-19. 
     \subsubsection{Publication repositories} These repositories shortlist the global scientific publications specific to COVID-19. The standard repositories include WHO COVID publication repository~\cite{WHO_Repo}, NIST COVID-19 repository~\cite{NIST_Repo}, CDEI COVID-19 repository and public attitudes retrospective~\cite{CDEI_Repo}, and the Elsevier Coronavirus Research Repository~\cite{Elsevier_Repo}.  
    
     \subsubsection{General data repositories} These repositories, such as NIH Open-Access Data and Computational Resources~\cite{NIH_Repo}, Google Health COVID-19 Open Data Repository~\cite{Google_Repo}, iReceptor Repository~\cite{iReceptor_Repo}, Stanford Research Repository~\cite{Stanford_Repo}, Humdata~\cite{Humdata_Repo}, etc., house the genomic, biomedical, clinical, sensor and behavioral data collected on COVID-19. 
    
    \subsubsection{Epidemiological information repositories} These databases record and visualize daily infected cases, deaths, etc., in the US and the world. Some examples include Johns Hopkins COVID resource center~\cite{JHU_Repo}, Our World in Data~\cite{OwD_Repo} and CDC COVID Tracker~\cite{CDC_Repo}, etc.  
    
    \subsubsection{Resource repository} These repositories help organize resources necessary for pandemic management. For example, Faith and COVID-19: Resource Repository~\cite{Faith_Repo} offers a web platform that allow people to participate in COVID-19 response, WASH Resources~\cite{WASH_Repo} offer intellectual resources to the staff and practitioners of Water Sanitation and Hygiene, Sages COVID-19 Medical Device Repository~\cite{Sages_Repo} offers information on medical toolkit for treatment of COVID-19, and COVID-19 Toolkit: Federal Depository Library Program~\cite{Fdlp_Repo} provides information on library closures, virtual work environments, and virtual service environments. 

\begin{figure}[h!]
	\centering
	\includegraphics[width=3in]{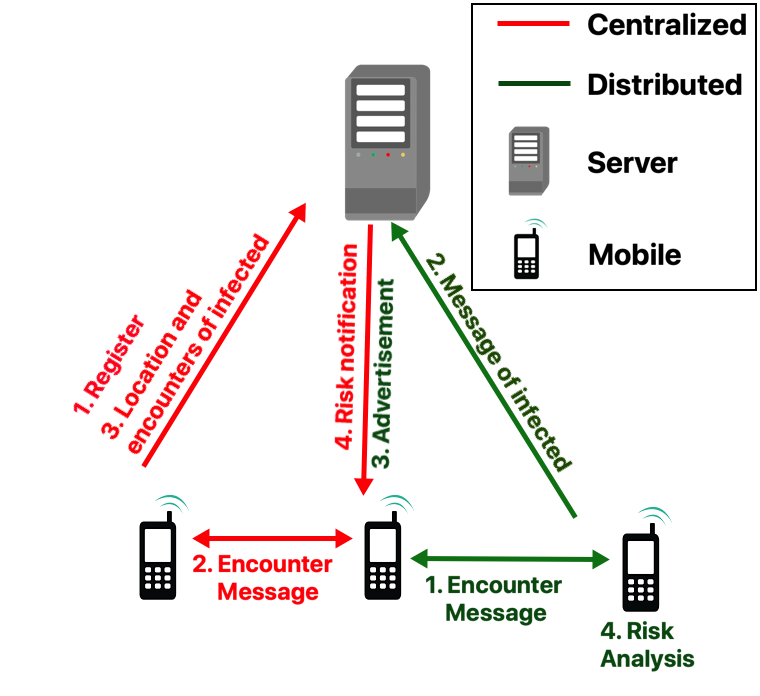}
	\caption{Message exchanges between server and mobile devices in the centralized and distributed architectures of mobile contact tracing applications}
	\label{fig:apps}
\end{figure}

\subsection{Mobile Contact Tracing Applications}\label{sec:contact}

\noindent Let us briefly discuss the two design architectures of contact tracing applications~\cite{ahmed2020survey} (see Fig. \ref{fig:apps}).

\subsubsection{Centralized architectures}
These applications, like Singapore’s TraceTogether~\cite{liang2020covid19} and China’s Health Code~\cite{dar2020applicability}, rely on the centralized back-end server to process the contact tracing information. The server carries out the authentication, risk analysis as well as user notification. We outline the four salient processes in the workings of centralized architecture-based apps (marked red in Fig. \ref{fig:apps}).

\begin{enumerate}
    \item The user registers their device with the server. User identity is authenticated by exchanging a one-time password. 
    
    \item The user app exchanges Bluetooth encounter messages with peers. The timestamp as well as the Received Signal Strength Indicator (RSSI) is \textit{locally} recorded. 
    
    \item If a user has tested positive and has agreed to share their medical status, the location information as well as the list of recent encounters are offloaded to the server.
    
    \item The server estimates the risk of exposure of the peers who came in contact with the infected person. The assessment is based on RSSI and transmission power. Subsequently, the likely exposed individuals are notified and prompted to isolate and get tested.
\end{enumerate}


\subsubsection{Distributed architectures}
These architectures, like Google-Apple contact tracing apps~\cite{michael2020behind}, do away with the dependency on the server, by performing more computational tasks on the user end. We discuss the key steps in distributed architecture based applications (marked green in Fig. \ref{fig:apps}).

\begin{enumerate}
    \item The devices do not register with the server nodes. They periodically exchange short messages, comprising pseudo-random key and current time, with the peers. 
    
    \item If a person tests positive, (s)he uploads these messages (and not the list of recent encounters) to the server. 
    
    \item The server advertises the message from the infected user to all mobile users.
    
    \item Users accessing the advertised messages do risk analysis locally based on proximity and duration of exposure. 
\end{enumerate}

\subsection{Vaccine distribution strategies}
These approaches address (1) equity and fairness and (2) epidemiological goals, such as minimization of infection, hospitalization and mortality in the course of vaccine allocation. Emanuel et al. discussed the three fundamental values of vaccine distribution, namely, maximizing people and limiting harm, prioritizing the disadvantaged, and equal moral concern~\cite{liu2020multivalue,emanuel2020ethical}. Overall, it is imperative that the vaccine distribution is commensurate with the population size to help achieve aforementioned epidemiological goals~\cite{matrajt2021vaccine}.  

\subsubsection{Optimization formulation}\label{sec:vac_opt} The standard approach to design vaccine distribution solutions is to model it as an optimization problem with constraints on the supply of vaccines. Roy et al. designed a generalizable, multi-vaccine distribution measure that allocates vaccines based on the socio-economic, epidemiological and demographic profiles of zones~\cite{roy2021optimal,roy2021generalizable}. In this approach, there is a matrix of decision variables $x_{w, b}^v \in \mathbf{X}$ (where $x \in [0, 1]$) that represents the fraction of vaccines of type $v$ transported from the vaccine warehouse $w$ to zone $b$. Optimization allocates $|Z|$ vaccines among $b \in B$ zones, while minimizing the cost, denoted by $d(w, b)$, between the warehouse $w$ and zone $b$ (Expression. \ref{eq:obj}). 

\begin{align}
    & \min_\mathbf{X} \sum_{v \in V} \sum_{w \in W} \sum_{b \in B} x_{w, b}^v \times d(w, b) \label{eq:obj}\; \\ 
    & \text{s.t.} \hspace{3mm} \text{C1:} \sum_{v \in V} \sum_{w \in W} \sum_{b \in B} x_{w, b}^v \leq |Z| \label{eq:vac}\; \\
    & \hspace{8.5mm} \text{C2:} \sum_{v \in V} \sum_{w \in W} x_{w, b}^v \geq c \times \tau \times |Z| \hspace{3mm} \forall b \in B \label{eq:con}\; \\  
    & \hspace{9mm} x_{w, b}^v \in [0, 1]
\end{align}

\paragraph{Number of vaccinations}  
\noindent The formulation allows the optimizer to distribute $|Z|$ vaccines, where inequality is replaced by an equality in Constraint 1 (C1, Ineq. \ref{eq:vac}). It preserves the inequality sign ($\leq$) if the goal is to minimize the number of distributed vaccines while meeting the following goals.  

\paragraph{Fairness based on demographic and epidemiological considerations}\label{sec:fair}
The optimizer can allocate vaccines based on the susceptible count, infected count or population density depending on the value of $c$ in Constraint 2 (C2, Ineq. \ref{eq:con}).  

    \begin{enumerate}

        \item \textit{Susceptible.} Vaccines are allocated to zones based on the proportion of remaining susceptible population (after $r_v \times \sum_w X_{w, b}^v$ have acquired immunity).
        
        \begin{equation}
            c = \frac{S_b - \sum_{v\in V} r_v \times \sum_w X_{w, b}^v}{\sum_{b \in B} S_b}
        \end{equation}
        
        \item \textit{Population Density.} Vaccines are assigned to zones based on the population density of each zone $b \in B$. 
        
        \begin{equation}
            c = c \times \delta_b
        \end{equation}
        
        \noindent In the above equation, $\delta_b = \frac{D_b - \mu(D)}{\sum_{b \in B}D_b}$, where $D_b$ is the population density of zone $b$.

        \item \textit{Infected.} Vaccines are allocated based on the ratio of infected population for each zone $b \in B$. 
        
        \begin{equation}
            c = c \times \iota_b
        \end{equation}
        
        \noindent Here $\iota_b = \frac{\rho_b - \mu(\rho)}{\sum_{b \in B} \rho_b}$, where $\rho_b$ is the infected ratio of zone $b$ given by $\frac{I_b}{S_b}$. 
    \end{enumerate}

\noindent Here, $\delta_b$ and $\iota_b$ denote the scaled population density and infected proportions for a zone $b$. If $\delta_b$ (or $\iota_b$) is greater than 1, its population density $D_b$ exceeds mean density $\mu(D)$, and vice versa. Summing over all $\delta_b$ (or $\iota_b$), i.e., $\sum_b \delta_b = \sum_b \frac{D_b - \mu(D)}{\sum_{b \in B}D_b} = \sum_b \frac{D_b}{\sum_b D_b} - \sum_b \frac{\mu(D)}{\sum_b D_b} = 1 - 1 = 0$. They consider a trade-off parameter $\tau \in [0, 1]$ balancing the economics vs. demographic or epidemiological factors. Specifically, a high $\tau$ distributes vaccines on the basis of the factors discussed under Constraint 2; conversely a low value of $\tau$ causes the optimizer to emphasize on economic cost.

There are several works that utilize a similar optimization formulation that distributes vaccines among communities based on their direct and herd immunity and mortality minimization~\cite{jadidi2020targeted,buhat2021using,matrajt2021optimizing}. Bertsimas et al. proposed a model that combines DELPHI epidemiological model with a bilinear, non-convex optimization to minimize exposure, mortality, and distance between vaccination centers and population centers~\cite{bertsimas2021locate}. Rao et al. present an analytical model for vaccine allocation with the aim of minimizing the effective reproduction number~\cite{rao2021optimal}. Shim et al. learn the parameters of the age-structured model of COVID-19 spread in South Korea to design vaccine deployment policies such that infections, deaths could be mitigated, given the parameters of age distribution and social contact~\cite{shim2021optimal}, while Meehan et al.~\cite{meehan2020age} and Buckner et al.~\cite{buckner2021dynamic} use the age-structured model and epidemiological model, respectively, to suss out high-priority individuals to be vaccinated based on their contact rates and risk of infection.  

\subsubsection{Measures for vaccine distribution} Jadidi proposed using bluetooth on mobile devices to create a social graph that will identify the most well-connected individuals to be vaccinated to achieve herd immunity~\cite{jadidi2020targeted}. These individuals, $u$, in a social graph have a high connectivity centrality measured as:

\begin{equation}
    C_{con} (u) = \sum_v c(u, v)
\end{equation}

\noindent Given the weight of the social tie between individuals $u$ and $v$, $w (u, v)$ measuring the likelihood of disease transmissibility based on the duration and frequency of contacts between the people, the distance, and their locations; and the number of simple paths between nodes $u$ and $v$, $h (u, v)$, $c(u, v) = \frac{w(u, v)}{h(u, v}$.

Xu et al. presented a vaccination coverage metric that determines the percentage of vaccinated individuals~\cite{xu2021hub}. The proposed metric improves upon the transportation accessibility and demographics constraints of the existing coverage metrics. Different zones are assigned levels $k = 1, 2, \cdots$ depending on the increasing amount of tine needed to reach the vaccination site of the zone from the neighborhood (refer to Sec. \ref{sec:dist}). The measure for zones in level $k$ to be covered by site $i$ is:

\begin{equation}
    \alpha_{i, k}^{*} = \min (1, \alpha_k + b \times N_i)
\end{equation}

\noindent In the above equation, $\alpha_k$ is the baseline of coverage for level $k$ zones as per Lim’s study~\cite{lim2016coverage}, $N_i$ is the total number of public transportation stops  in the vicinity of site $i$, and $b$ measures the convenience rendered by adding a stop near a given site $i$. 

\subsection{Dynamic lockdown strategies} We discuss the models that prescribe varying levels of lockdown based on the epidemiological and clinical parameters, such as infected cases, mortality, contact rates, healthcare facilities, etc., of zones constituting a geographical region.

\subsubsection{Supervised models} Tadano et al. employed artificial neural networks (namely, extreme learning machine, echo state network; multilayer perceptron, and radial basis function networks) to predict the effect of lockdown in Sao Paulo City, South America on air pollution levels~\cite{tadano2021dynamic}. Roy et al. propose a dynamic pandemic lockdown strategy that leverages reinforcement learning and queueing models to regulate inter-zone traffic on the basis of hospitalization and healthcare budget~\cite{roy2021towards}. Roux et al. proposed an age-structured Susceptible-Exposed-Infected-Recovered model (refer to Sec. \ref{sec:seird}) to estimate the right number of hospitalizations, hospital beds requirements, and hospital deaths that could have been prevented by a timely lockdown in 13 metros of France~\cite{roux2020covid}. They employed maximum likelihood estimation to train the model and show that without travel restrictions, the pandemic would have overwhelmed the medical facilities in France.

\begin{figure}[h!]
	\centering
	\includegraphics[width=3in]{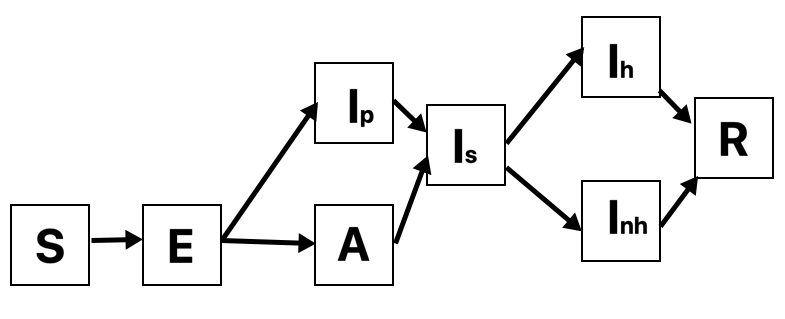}
	\caption{SEIR model proposed by Roux et al.~\cite{roux2020covid}, where $S$, $E$, $I_p$, $A$, $I_s$, $I_h$, $I_{nh}$, and $R$ represent susceptible, exposed, pre-infected, asymptomatic, symptomatic infected, hospitalized, non-hospitalized and recovered, respectively.}
	\label{fig:seir}
\end{figure}

\subsubsection{Unsupervised models} Rahman et al. proposed an unsupervised machine learning based dynamic lockdown framework. The clustering algorithm identifies dynamic clusters based on the infected cases and mobility data. The clusters are marked for hard, moderate or no lockdown depending on the number of active or suspected clusters in them~\cite{rahman2020data}. Gollier incorporated the uncertainty of the infection status of an individual into the SIR epidemic model to infer the economic impact of the lockdown imposed~\cite{gollier2020pandemic}. Similar to the study by Tadano~\cite{tadano2021dynamic}, Bao et al. performed regression analysis on 44 cities in northern China to show the strong association between reduction in air pollution and strong lockdown~\cite{bao2020does}. 

Bosi et al. challenge the notion of indefinite immunity of the recovered (or immunized) individuals~\cite{bosi2021optimal}. They employ a dynamic equilibrium model to show that positive lockdown is useful when the society is altruistic (i.e., willing to maximize collective rewards), while zero lockdown is an optimal lockdown policy if the agents are selfish. Pichler et al. propose a dynamic input-output model to gauge the ramifications of COVID-19 and its consequent lockdown on the demand and supply chain of industries in the UK~\cite{pichler2021and}. Fajgelbaum et al. proposed an optimization to solve the planning problem of determining the traffic flow across districts in two Korean cities that reduce the economic overheads due to uncoordinated lockdown measures as well as the loss of lives~\cite{fajgelbaum2020optimal}. The social planner optimization formulation works as follows:

\begin{equation}
    W = \max_{\chi (t)} \int_0^{\infty} e^{-(r + v)t} \sum_t [U (j, t) + \frac{v}{r} \times _U^{\sim}  (j, t) - \omega \gamma_D I (j, t)] dt   
\end{equation}

\noindent This equation maximizes the discounted value of real net income considering loss of lives, where $U (j, t)$ is the real income of location $j$ at time $t$ based on the distribution of infected cases and lockdown and $_U^{\sim}$ is the income if the vaccine is freely available with probability $v$. The variable $\omega$ represents the expected income of the COVID victims (measured as their expected lifetime times annual income minus the discounted value of wages) and $I (j, t)$ is the daily infected at location $j$ at time $t$. 

Askitas et al. proposed a multi-event model to study the effects of different overlapping and non-overlapping intervention measures varying in timing, duration, and intensity on the overall incidence of COVID-19 and the resultant mobility patterns~\cite{askitas2020lockdown}. Pestieau studied how the lockdown strategies, namely uniform lockdowns and age-differentiated lockdowns, align with social welfare criterion, balancing the trade-off between economy and lives lost~\cite{pestieau2022optimal}. Finally, Charpentier et al. extended the SIR epidemic model to propose the optimal lockdown model that minimizes the socioeconomic impact of the pandemic, while meeting the capacity of the ICU~\cite{charpentier2020covid}.

\subsection{Contact tracing and mobility scheduling} \label{sec:schedule}

The contact-tracing applications discussed in Sec. \ref{sec:contact} already inform human mobility~\cite{petrovic2021data}. While they are yet to get traction in the realm of pandemic management, the apps possessing simple user interface and visual depictions of infection hotspots is truly the future of pandemic management~\cite{heikkila2021planning}. Ranisch et al. pointed out that several ethical and humanitarian challenges, such as, data protection, potential stigmatization of patients, social justice concerns, etc., need to be ironed out before these apps get universal acceptance~\cite{ranisch2020digital}. These apps are connected to smart mobile devices~\cite{serafino2021superspreading} or embedded in wearables~\cite{tripathy2020easyband}. We now discuss the models some of which have been implemented as mobile applications to recommend destinations so that contagion can be minimized. 

Hu et al. delineated the human mobility data sources and computational approaches that aid computational research on pandemic management. They compared the attributes of these datasets on the basis of privacy, quality of information, storage, processing and access~\cite{hu2021human}. Similarly, Grantz discussed the applicability of mobile applications in transmission and control of contagion as well as the implication of selection bias in mobile phone data~\cite{grantz2020use}. Oliver et al. delve into some of the shortcomings of such data-driven research and ways to overcome them through the involvement of government and domain experts~\cite{oliver2020mobile}. Benita et al. apply text mining to find themes in scientific literature on human mobility behavior, while reporting keywords and future research directions~\cite{benita2021human}. 

Based on GPS-based mobility data from US counties, Tokey employed space-time cube, curve-fitting and regression to study the evolving relationship between human mobility and COVID daily infected cases over time with  alteration in socioeconomic, demographic and geographic policies~\cite{tokey2021spatial}. Loo et al. utilized mobility data from public places, like bars, shopping centres, karaoke/cinemas, mega shopping malls, public libraries, and sports centres, to learn the connectivity among populous zones and their potential to serve as infection hotspots~\cite{loo2021identification}. Analogously, Orallo \& Martinez explored the applicability of pure synthetic and simulated mobility models on predicting the risk of contagion~\cite{hernandez2021human}. Basu discussed how increasing dependency on private vehicles and dwindling mass transit ridership is a hindrance to affordable and sustainable urban mobility in the post-COVID world~\cite{basu2021sustainable}.

\begin{figure}[h!]
	\centering
	\includegraphics[width=3in]{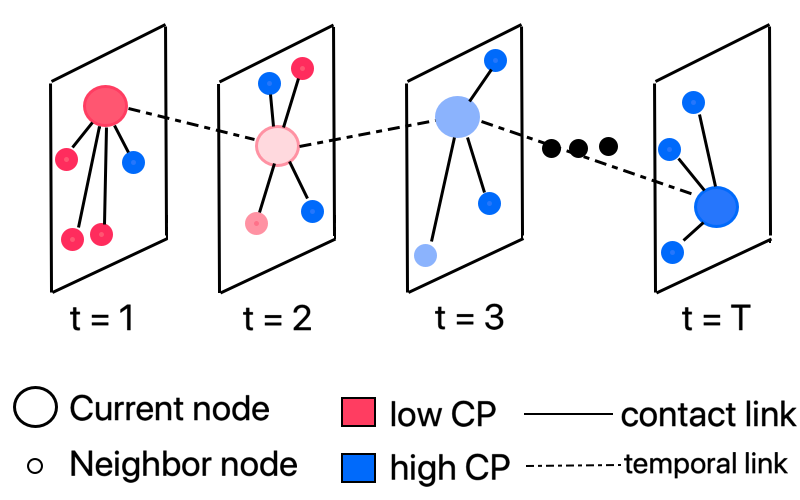}
	\caption{Evolution of contagion potential (CP) over time. Every panel shows the location of nodes at time $t$ ($1 \leq t \leq T$) and the spectrum of colors - dark red, light red, light blue and dark blue - represent increasing CP of the observed node (large circle) on contact with infected individuals (small circles).}
	\label{fig:cp}
\end{figure}

Roy et al. leveraged network science to propose three optimization strategies, which will result in social networks that minimize contact between the infected and susceptible individuals~\cite{roy2021leveraging}. As part of the same goal, they introduced the notion of contagion potential that measures, on a scale of 0 to 1, the likelihood of an individuals to act as spreaders of contagion (CP). Instantaneous CP ($P_t(u)$) of an individual is measured as the number of infected people (s)he comes in contact with at time $t$ (see Fig. \ref{fig:cp}). The overall CP till time $T$ is calculated as the mean instantaneous value CP, as follows: 

\[
    Z_t(u) = 
\begin{cases}
    0, & u \in R, D\\
    1, & u \in I\\
    \frac{1}{T} \sum_{t = 0}^T P_t(u) & \text{Otherwise}
\end{cases}
\]

\noindent They introduced a mobile app, called \textit{MyCovid}, that includes a case study on the proposed optimization strategies and guide the users’ mobility to minimize contagion. With prior user permission, the app creates a repository of mobility traces and enables research on informing mobility during outbreaks.

Serafino et al. explored the GPS traces of mobile devices in Latin America to infer a social contact network, based on the duration of contact and proximity between the infected and susceptible individuals~\cite{serafino2021superspreading}. They demonstrated that the lockdown reduced human mobility, but did not impede disease spread. They employ percolation theory to explain this idea. In bond percolation, the network connectivity is reduced by removing a small fraction of links (bonds) in the contact network. Serafino measure network connectivity in terms of the size of the giant connected component (GCC) (calculated as the largest set of nodes such that each node is reachable from every other node in that list). Through network analysis, they depict that the lockdown measures causes a drop in GCC, thereby causing the dense contact network to be reduced to several modules of strongly connected modules. 

\begin{figure}[h!]
	\centering
	\includegraphics[width=2.5in]{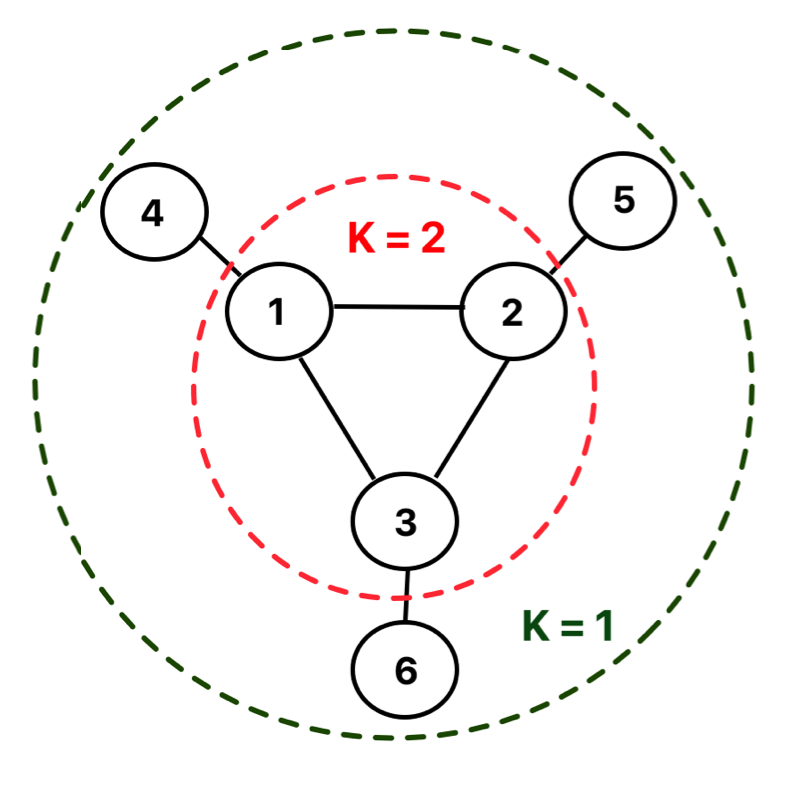}
	\caption{A 6-node network with $K = 2$, where 2-core and 1-core nodes are marked red and green, respectively}
	\label{fig:Kc}
\end{figure}

\noindent However, despite the drop in GCC, the number of cases continued to grow, albeit at a lower pace. Serafino used the notion of $K-$ core~\cite{dorogovtsev2006k} to explain this phenomenon. $K-$ core is obtained by iteratively pruning the lowest degree nodes from the network (see Fig. \ref{fig:Kc}). The high K-cores possess a high disease transmission persistence. Despite the absence of the
GCC, the maximal k-core remained in the contact network, resulting in growing cases. Similarly, Roy et al. also employed machine learning and network science concepts, like coloring and clustering, on daily infected cases in New York City to propose interdiction strategies that mitigate contagion through restricted human mobility, while meeting economic goals~\cite{roy2021effectiveness}.

\subsection{Pandemic Prediction and Travel Policy}\label{sec:surv_pand} 
\begin{figure}[h!]
	\centering
	\includegraphics[width=3.5in]{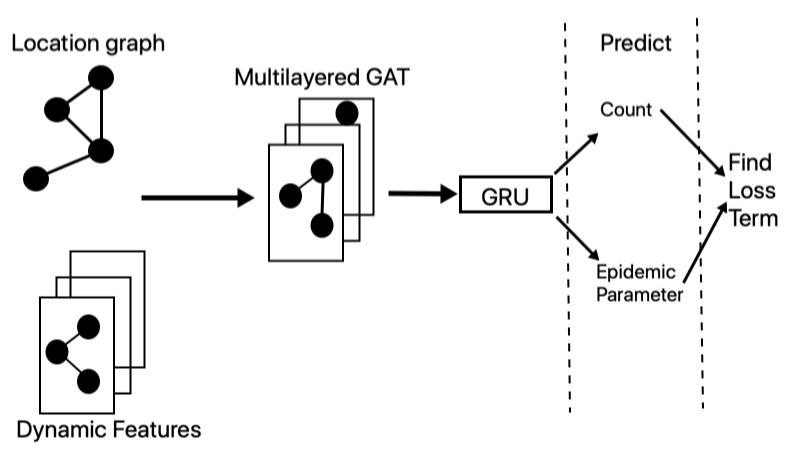}
	\caption{Spatiotemporal prediction architecture~\cite{gao2021stan}, which employs graph attention network (GAT) and gated recurrent unit (GRU) to predict future trends in infection and recovery.}
	\label{fig:stan}
\end{figure}

In Sec. \ref{sec:schedule}, we discussed mobility scheduling techniques which employ computational models on human mobility data to inform short-distance mobility. The current section is dedicated to a survey on efforts to make spatiotemporal prediction that determine global pandemic-related travel policy. Gao et al. proposed a spatiotemporal attention network (STAN) approach for pandemic prediction~\cite{gao2021stan}. The proposed framework leverages static features (namely, geographic and demographic data of states and counties) and dynamic features (namely, daily infected counts). Fig. \ref{fig:stan} shows that the framework constructs locations graphs with zones as nodes and inter-zone interaction as links for the static and dynamic features.

\noindent STAN utilizes a graph attention network (GAT) to learn the spatiotemporal interaction across zones. GAT network employs a multihead approach to learn attention scores denoting the interaction between any pair of zones (refer to Sec. \ref{sec:graph} for summary on graph theory) and combines the attention scores to infer the embedded representation of the zones. Subsequently, STAN processes the graph embedding with a gated recurrent unit (GRU) -- a type of recurrent neural network -- to predict the future epidemiological parameters and infection and recovery counts and update the loss terms. Gao et al. discuss the applicability of STAN in building spatiotemporal models for disease status and healthcare resource utilization~\cite{gao2021stan}.

Pinter et al. introduced a hybrid model that combines network-based fuzzy inference system and multi-layered perceptron-imperialist competitive algorithm to predict the infected and death counts in Hungary~\cite{pinter2020covid}. Kumar et al. used PROPHET and ARIMA disease forecasting models to gauge the trends in contagion in US, Spain, Italy, France, Germany, Russia, Iran, United Kingdom, Turkey, and India~\cite{kumar2020covid}. Singhal et al. also proposed a mathematical series-based model and Fourier decomposition method to predict trends in future cases in India, Italy, and USA~\cite{singhal2020modeling}. Chowdhury et al. introduced a multivariate model that uses epidemiological and clinical parameters to forecast the spread trajectory in 16 countries~\cite{chowdhury2020dynamic}. 

Like Kumar, Oguntokun et al. and Alzahrani et al. employed the ARIMA model to evaluate the possibility of air- and water-based transmission of the pathogen and the impact of preventive measures on contagion in India and Saudi Arabia, respectively~\cite{ogundokun2020machine,alzahrani2020forecasting}. Kristjanpoller et al. presented frameworks that exploited meta-learners and causal forest models to achieve dynamic quarantine policy and model intervention during future outbreaks. They validated the dynamic policy on the epidemiological data of Chile because of its high infection numbers and socioeconomic and demographic heterogeneity~\cite{kristjanpoller2021causal}. Finally, Luo discussed the variables influencing the accuracy of the time-series prediction models, namely, new strains, human behavior, politics, etc., and emphasized on the need for predictive monitoring to better understand their efficacy~\cite{luo2021forecasting}.

Watson et al. used Bayesian time series model and a random forest algorithm to predict infected cases and death numbers. Specifically, the Bayesian model is trained on the cumulative case count to project distribution of future cases, while the random forest algorithm works in tandem with the epidemiological model to predict pandemic deaths~\cite{watson2021pandemic}. Roy et al. introduce a spatiotemporal network-inference approach that uses a sliding window to scan daily infection numbers. The resultant temporal networks quantify the potential of inter-zone infection spread and help identify zones with a high outflow of link weights as ones acting as disease hotspots~\cite{roy2021spatiotemporal}. Zhang et al. introduced a minimization formulation that determines the extent of intervention over a period of $J$ days in terms of the population size $N$~\cite{zhang2020effect}, measured as:

\begin{equation}
    \min_N \frac{1}{J} \sum_{j = 1}^{J} e_j (N)
\end{equation}

\noindent In the above equation, $e_j (N)$ denotes the prediction error in cumulative infection for a given population count $N$.

Mehta et al. implemented a machine learning approach on fused health statistics, demographics, and geographical datasets to gauge the course of the pandemic. They predicted the risk of outbreaks in the US counties on the basis of their urban and vulnerable population demography~\cite{mehta2020early}. Ahmad et al. discussed the role of big data analytics and IoT in designing a neural network-based health monitoring framework that possess  diagnostic, predictive, and prescriptive capabilities to predict the future of the pandemic~\cite{ahmed2021framework}. Similarly, Kavadi et al. presented a Nonlinear Global Pandemic Machine Learning model for global pandemic prediction. They applied the model on demographic, clinical and epidemiological data of Indian states to show its efficacy in pandemic prediction~\cite{kavadi2020partial}.

\section{Pandemic Management Architecture}\label{sec:arc}

\subsection{Architecture Overview} 

We propose a three-layer architecture for managing future pandemic situations. Currently, different federal governments' systems for the COVID19 pandemic management are ad-hoc and loosely coupled. These systems lack a proper and coherent ecosystem needed for dynamic decision-making. Our envisioned ecosystem will gather sensory data from citizens (such as co-location information, the health status of the individual, and so on.), share them dynamically over the network to remote servers, fuse the ground data with the historical time-series epidemiological data fed from various other sources at the servers, and sends messages to civic authorities as well as users to take appropriate measures. 

\begin{figure*}[h!]
	\centering
	\includegraphics[width=5.75in]{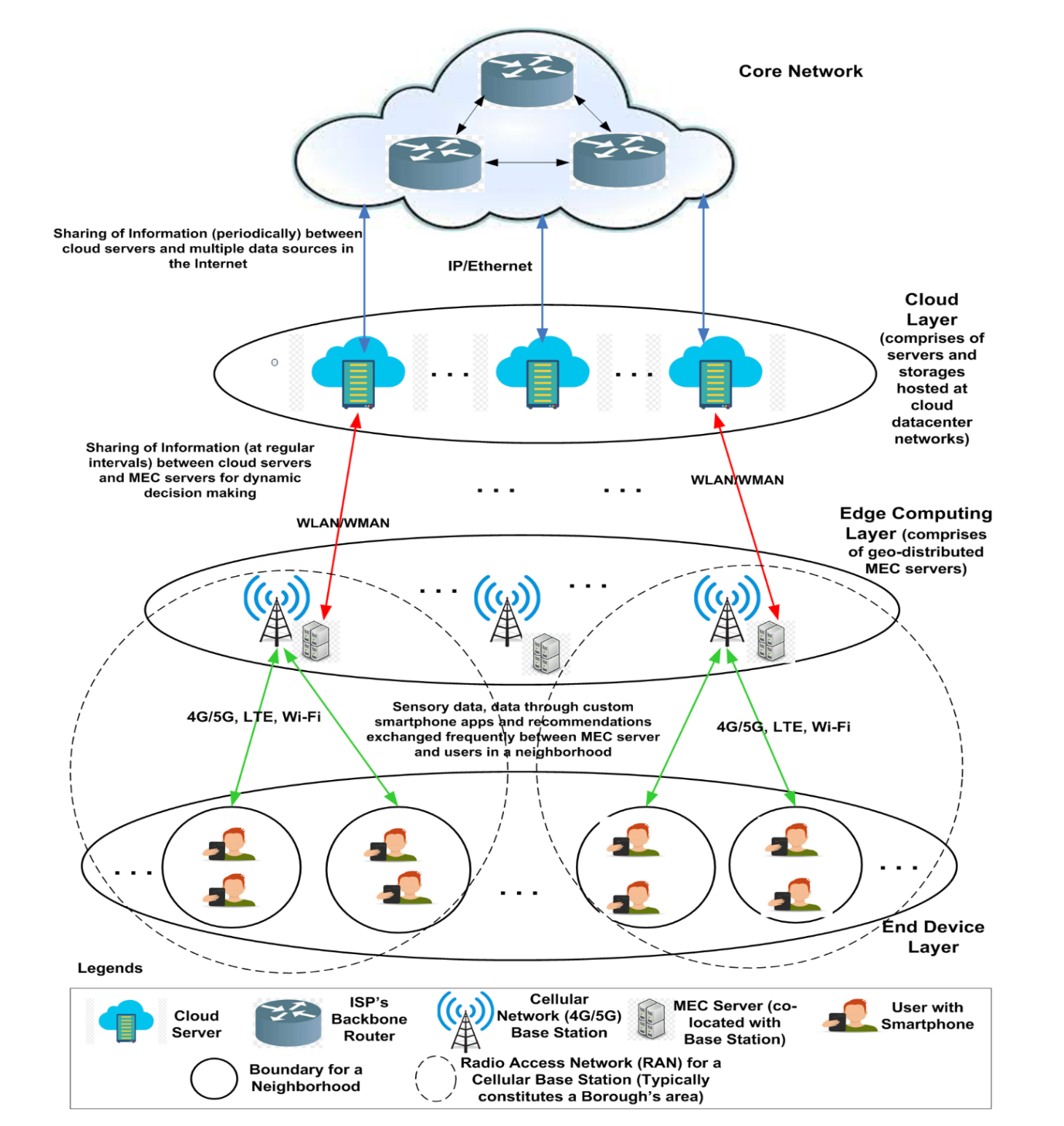}
	\caption{The pandemic management architecture comprising the cloud layer, edge computing layer, and mobile users.}
	\label{fig:arc}
\end{figure*}

A schematic representation of our envisioned architecture is given in Fig. \ref{fig:arc}. It consists of three distinct layers – (1) cloud layer; (2) edge computing layer; and (3) end device layer. We propose using the edge computing paradigm to provide networking support for collecting and sharing data across the different layers. The objective of edge computing is to introduce an intermediate computing layer between the end device (bottom) and cloud layers (top) such that compute, storage, and communication services can be brought to the edge of the end devices' network. Another motivation behind using edge computing is to handle the massive scale of the end devices and the petabytes of digital footprint they create nowadays. Offloading sensory data onto remote servers creates bandwidth bottlenecks in the backhaul networks, increasing delay and performance loss. Thus, if the sensory data are delegated to edge nodes that are in proximity to the network edge, not only bandwidth wastage in backhaul networks can be avoided, but also response time, throughput, and latency can be improved to facilitate a better user experience. We describe the different components of this architecture below. 

\subsubsection{Core network}
The core network (Internet), managed by several regional ISPs, hosts a plethora of epidemiological, healthcare, and demographic data hosted by different sources. These sources include Worldometer, CDC-USA, Johns Hopkins CSSE, and Ministries of Health or other Government Institutions and Government authorities' social media accounts. Some of the features of such data are – population count, population density, number of cases (infection count), mortality rate, R-factor, number of vaccine doses administered, number of tests performed, the spread of infection in terms of count per ten thousand/hundred thousand/million and so on. Also, healthcare data on the number of hospitals, number of beds in the COVID ward per hospital, availability status of COVID19 related medicines, etc., are made available in these data stores. All data can be of different granularities depending on the macro-level or micro-level views. Most of them are time-series data updated tentatively after every 24 hours. 

\subsubsection{Cloud layer} \label{subsubsec:cloud_layer}

This layer consists of an array of high-performance servers hosted in the cloud capable of handling large data volumes and carrying out massive computation tasks in parallel. The cloud-hosted servers will periodically (maybe once in a few hours) pull data from different sources and maintain an updated repository. The servers use IPv4/IPv6 routing as the backbone network technology to pull data from the public/paid APIs provided by various sources on the Internet. This data is used by the following four significant modules (discussed later in Sec. \ref{sec:edge_comp}) running in these servers:

\begin{itemize}
\item \textit{Vaccine allocation}: This module will find optimal vaccine allocation strategies across different zones of geographical space (city, county/district, state, etc.). 

\item \textit{Dynamic lockdown}: This module is responsible for determining if any zone requires lockdown (partial or complete) based on its current healthcare facilities available, infection spread, mobility of people, and other factors. 

\item \textit{Contact tracing and mobility scheduling}: The purpose of this module is to find people who are close to an infected individual. If any zone is found to have many unvaccinated and exposed individuals, then a spatiotemporal schedule can be recommended to curb the uncontrolled movement of large masses across the zone. 

\item \textit{Pandemic prediction}: In this module, one or more prediction algorithms will be running to infer the likelihood of an outbreak of the pandemic situation at a particular zone. The algorithm will use healthcare infrastructure data, all types of epidemiological data, and mobility and contact tracing information for making the prediction. Such predictions will enable decision-makers to allocate additional resources to that zone and prevent an unforeseen health crisis owing to the pandemic. 
\end{itemize}

\subsubsection{Edge computing layer}

The edge computing layer consists of a network of mobile edge computing (MEC) servers~\cite{hu2015mobile}, each co-located with a cellular base station. MEC servers are managed by different mobile network operators and have heterogeneous system resources (CPU, memory, storage, bandwidth). A MEC server uses the radio access networks (RANs) of its co-located base station to collect sensory data from citizens spread across multiple neighborhoods of a borough in an urban city. Therefore, several MEC servers will be deployed in different boroughs to collect as much human-sensed data as possible from different neighborhoods across the city. The pandemic management agencies need to roll out smartphone-based app(s) for collection of citizens' data who are at different neighborhoods. We assume this/these app(s) is/are installed by the users on their devices and necessary permissions are granted for data collection. 

The MEC servers can provide on-demand compute, cache, and communication services to data collection application(s) running on the user devices. The server instances for these applications run on the MEC servers. They are responsible for collecting information on vaccination status and epidemiological condition (i.e., susceptible, infected or recovered) from individual users and location (waypoint) information from the devices. The periodic waypoint information enables the server to trace out the mobility of the users and also the places they are making visits. The past application data collected from the users is pushed to the back-end cloud servers periodically. This is useful in situations when the mobile users migrate to a new neighborhood beyond the control of its current MEC. The architecture handles this handoff by allowing the new MEC to download the pertinent information of that user.

The instances of the modules running on the cloud servers will also be hosted in the MEC servers to enable dynamic decision making at the edge of the users' networks. These computation modules will leverage the  sensory data collected from the public and fuse them with the historical ones (pulled from the back-end cloud storage) to develop recommendation(s) for the stakeholders (civic authorities, healthcare specialists, caregivers, public). However, if any MEC server is overloaded or unequipped with computational resources, the computation of the modules will be offloaded to the cloud. 

The application(s) hosted in MEC can send alerts to the individuals if they are currently present in a large gathering with a few infected people or if their health status necessitates isolation at the earliest. For communicating with cloud servers, the MEC servers will use Wi-Fi/WiMax as carrier technologies, while for gathering ground-level public data, wireless technologies such as Wi-Fi, LTE, 4G/5G, etc., will be used.  

\subsubsection{End device layer}
The end device layer consists of tens of hundreds of intelligent hand-held IoT devices such as smartphones, wearables, tablets, etc., possessed by citizens moving around an urban area. We collectively refer to them as user equipment (UE). These users, present in different neighborhoods and boroughs, are expected to install and register with the COVID19 app(s) and allow it/them to collect UEs' location information opportunistically. Besides this, the app(s) will also request the users to share information related to his/her vaccination status, current health condition, and previous health history (if any). The app(s) will push notifications from time to time, asking the user to update the information provided if required and also send notifications regarding the appropriate measures the individual has to undertake.

The UEs can either use Wi-Fi or LTE/4G/5G communication or a combination of the two (in hand-off scenarios to support mobility) to upload data to the application server(s) running on the MEC servers. To support dynamic decision making in presence of mobility, the proposed architecture should enable migration of relevant data if any user changes position from one RAN to the other. In such scenario, following the network hand-off, once the user signs back in to the app, his/her location is updated and corresponding epidemiological data will be migrated from cloud storage to the nearest MEC server for sending health advisory alerts and recommendations.\\ 

\noindent \textbf{Overview of the architecture.} Let us discuss a pandemic stack (see Fig. \ref{fig:stack}) that captures the logical flow of the proposed pandemic management architecture. The first layer is the \textit{application layer} that operates in two ways: (1) use of the mobile (crowdsensing and contact-tracing) applications (refer to Sec. \ref{sec:contact}) to collect individual clinical, epidemiological and physiological data and (2) the application programming interfaces to store and access data from the cloud servers. The next layer is the \textit{privacy (and security) layer} that ensures the data is obfuscated in a manner that the sensitive user information is hidden away, while maintaining data utility (refer to Sec. \ref{sec:priv}). Next, the \textit{storage layer}, comprising the data servers as well as local smart device memory, stores the anonymized data. As depicted in Fig. \ref{fig:arc}, the \textit{network layer} allows the interaction between (1) cloud and edge layers, (2) edge and mobile layers and (3) mobile users (see Fig. \ref{sec:future}) through WLAN, WMAN, LTE, 4G/5G, Wi-Fi, etc. technologies. Finally, there is the \textit{computational layer} that applies the computational models on the structured data to make prediction and recommendations over time.

\subsection{Details of Intelligent Computation Modules}\label{sec:edge_comp}
\noindent This section elucidates the workings of the four intelligent computation modules, the instances of which are running on both the MEC and the cloud servers. Fig. \ref{fig:sol} shows that the intelligent computation modules receive the data from the public repositories as well as the mobility, epidemiological and physiological data from the mobile users, and decides (in four separate modules) the vaccine distribution, dynamic lockdown, mobility schedules, and pandemic policy across the stipulated zones. Lastly, the effect of the four modules are reflected in the form of outcome variables recording the daily new infections, deaths, and the consequent pressure on the healthcare infrastructure of the zones. 

The computation modules have three features. First, the static and dynamic prediction model (see Sec. \ref{sec:pred}) that informs modules of the outcomes from earlier decisions and predicts future trends of inputs. Second, each module processes a predefined list of factors (refer Sec. \ref{sec:fact}) from the input data and the information from the prediction models. A module may apply several computational models separately and fuse their solutions using \textit{integration algorithms} (see Sec. \ref{sec:int}) to arrive upon decisions. Let us discuss each feature.

\subsubsection{Socioeconomic, demographic and epidemiological factors}\label{sec:fact} 
\begin{figure*}[h!]
	\centering
	\includegraphics[width=5.5in]{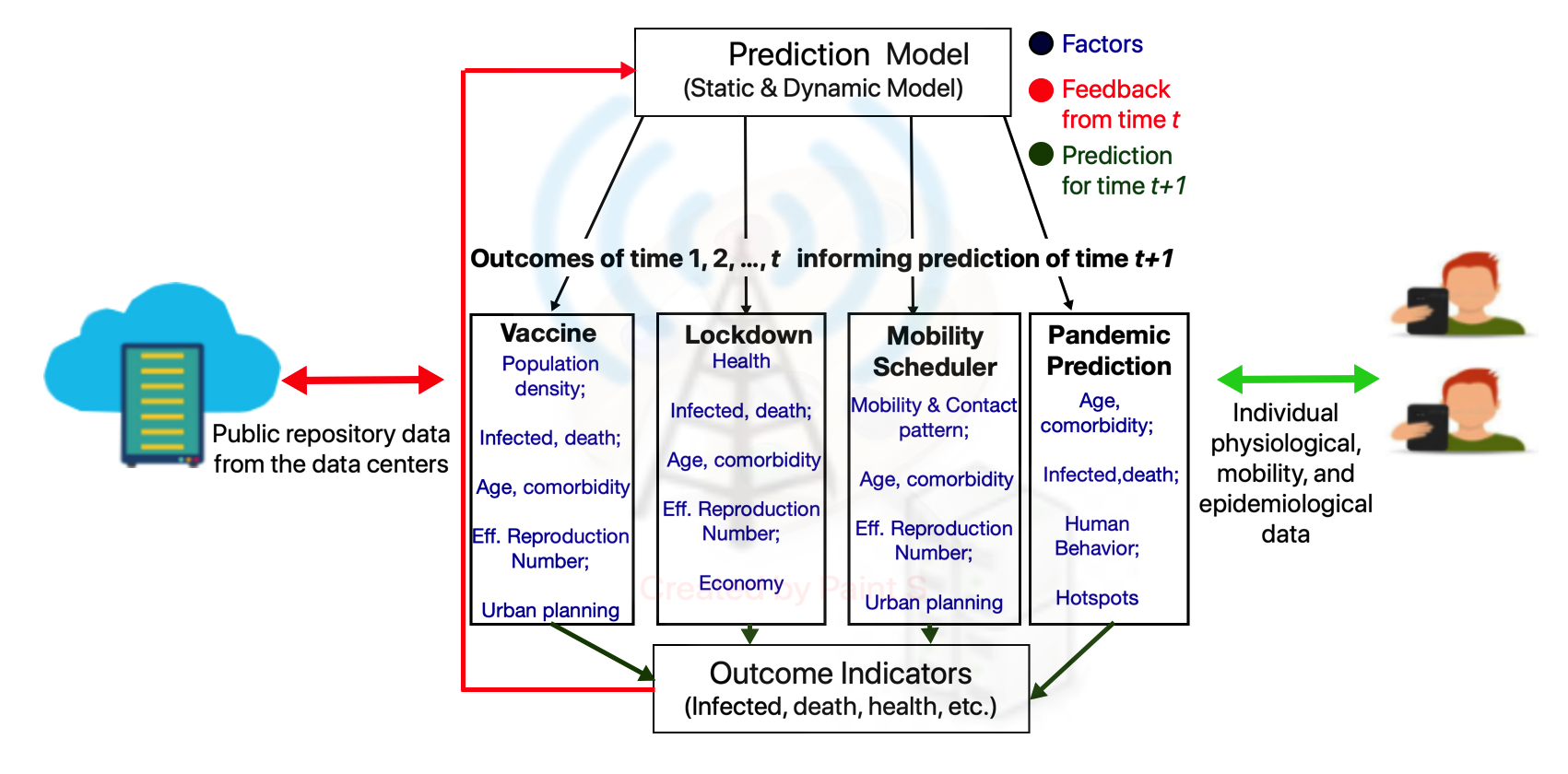}
	\caption{The computation module running vaccine distribution, dynamic lockdown, mobile scheduling and pandemic prediction modules on clinical, epidemiological and physiological data from users and public repositories to make recommendations.}
	\label{fig:sol}
\end{figure*}

We discuss the colored factors that serve as inputs (marked blue in Fig. \ref{fig:sol}) to the four modules. (In Sec. \ref{sec:survey}, we have discussed the computational studies on vaccine distribution, dynamic lockdown, mobility scheduling and pandemic prediction that utilize these factors to draw inferences.)

\begin{figure}[h!]
	\centering
	\includegraphics[width=3.25in]{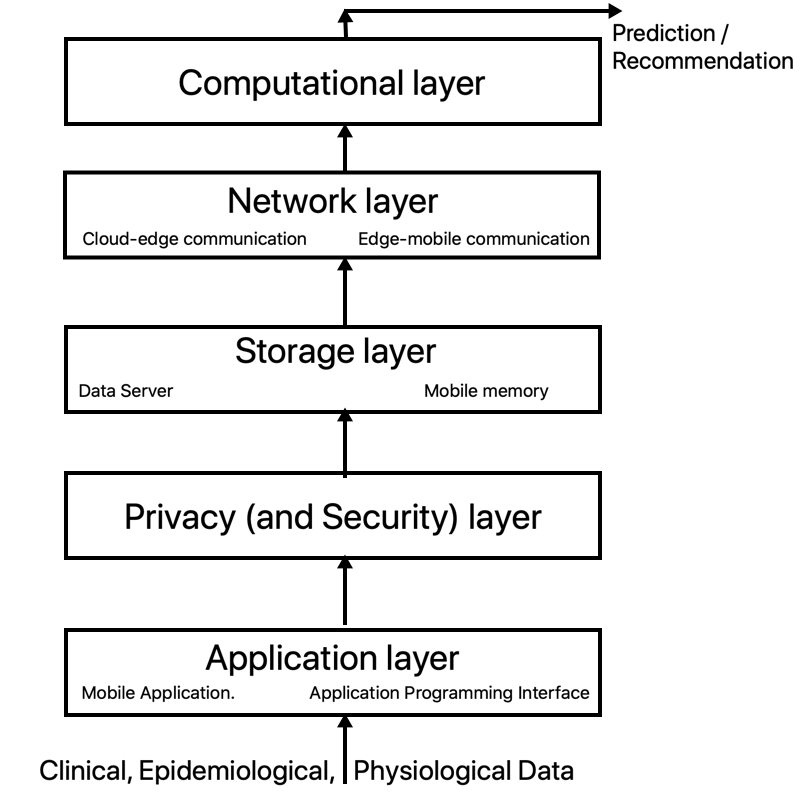}
	\caption{Logical flow of the pandemic stack}
	\label{fig:stack}
\end{figure}

\paragraph{Vaccine distribution} The vaccine distribution module determines the number of vaccines received by zones based on the following key factors (delineated during the survey in Sec. \ref{sec:vac_opt}). (This module can be extended to distribute drugs to treat symptoms of the virus.) Since \textit{population density} determines the frequency and duration of interaction among individuals, the vaccine distribution modules may attempt to distribute more vaccines to densely populated zones~\cite{roy2021optimal,buckner2021dynamic}. The module may distribute vaccines based on the disease transmissibility, gauged in terms of \textit{reproduction number} (see Sec. \ref{sec:seird}), among the zonal population~\cite{rao2021optimal}. The proportion of the \textit{elderly} individuals with \textit{comorbidities} may be a determinant of vaccine distribution~\cite{meehan2020age}. Poor \textit{socioeconomic condition} of the population or \textit{urban planning} of a zone may be grounds for the inoculation of its population ~\cite{lim2016coverage}.

\paragraph{Dynamic lockdown strategies} This module determines the duration and extent of lockdown to be imposed on zones depending on the changing socioeconomic and epidemiological conditions. The module may recommend the imposition of stricter lockdowns in zones with a poor \textit{healthcare infrastructure}~\cite{roux2020covid,roy2021towards} to prevent the patients from being denied treatment in the event of severe outbreaks~\cite{gaffney2020covid}. Since lockdowns have \textit{economic overheads}, the module must strike a balance between contagion and economy~\cite{gollier2020pandemic,pestieau2022optimal}. It may assess how the restrictions disrupt the supply-demand chain of industrial sectors~\cite{pichler2021and}. Finally, \textit{age and comorbidity} is another factor contributing towards lockdown recommendations.

\paragraph{Mobility scheduling} Based on the epidemiological data, this module informs individuals of their next location (via mobile applications) to protect them from disease contraction. We have discussed in Sec. \ref{sec:schedule} that \textit{mobility and contact pattern} of the zones is a key factor influencing the identification of hotspots~\cite{hernandez2021human,roy2021leveraging,dorogovtsev2006k,serafino2021superspreading}. Similarly, the knowledge of \textit{urban planning} can help determine the regions of high human contact~\cite{basu2021sustainable}. Also, \textit{age and medical history} are important considerations towards determining the level of priority in protecting individuals from exposure to infection.

\paragraph{Pandemic prediction} As discussed in Sec. \ref{sec:surv_pand}, this module considers bulk population mobility (and not individual mobility traces like the mobility schedule module) to determine future trends. \textit{Time-series infected and mortality numbers} of the population is essential to predict the course of the pandemic~\cite{gao2021stan,pinter2020covid}. \textit{Age and comorbidity distribution} of the zones help determine the extent and duration of the infection spread~\cite{mehta2020early}. \textit{Human behavior} is another factor influencing the prediction of these modules, since the level of adherence to lockdown restrictions affects how promptly contagion can be mitigated~\cite{luo2021forecasting}. Latest information on the \textit{number of hotspots and severity of outbreaks} in them is necessary in drawing up public policies related to the pandemic~\cite{roy2021spatiotemporal}.

\subsubsection{Data integration}\label{sec:int} For each module, several frameworks (as discussed in Sec. \ref{sec:survey}) can be applied to infer the respective decisions. Moreover, these frameworks may utilize varying subset of factors as inputs. For example, Rao et al. solve vaccine allocation based on effective reproduction number~\cite{rao2021optimal}, whereas Shim et al. address the same problem based on age and contact pattern~\cite{shim2021optimal}. Evidently, consensus algorithms are necessary to aggregate the solutions proffered by myriad computational frameworks. Monti et al. proposed an iterative sampling-based consensus algorithm, which uses a consensus matrix to assign cluster identity to each datapoint (i.e., solutions in our context)~\cite{monti2003consensus}. Subsequently, the consensus clustering strategies by Hoadley et al. attempt to reassign cluster labels to solutions based on vectors denoting solution clusters emerging from different approaches~\cite{hoadley2014multiplatform}. Shen et al. employed probabilistic matrix factorization~\cite{shen2009integrative}, while Wang et al. exploited network similarity properties for integration of heterogeneous data~\cite{wang2014similarity}. For every module, the edge device can adopt the above models to infer the aggregated decision.

\subsubsection{Prediction models}\label{sec:pred} There can be two types of prediction models: \textit{static} and \textit{dynamic}. The static prediction models could be \textit{supervised machine learning} models (namely, Naive Bayes, Decision Trees, Linear Regression, Support Vector Machines, Multilayer Perceptron, etc.) and \textit{unsupervised machine learning} models (clustering, associative rule mining, etc.)~\cite{berry2019supervised}. The dynamic models incorporate the time-series feature data from the past to predict the present outcomes~\cite{bontempi2012machine}. These models include univariate and multivariate time-series forecasting models, such as autoregressive integrated moving average, vector autoregressive models, long short-term memory, etc. Although both the static and dynamic models work on (1) past feature data $\mathbf{X}$ (comprising the factors) and the corresponding decisions from the four modules $\mathbf{D}$ and (2) observed outcome $\mathbf{Y}$ (i.e., infected, death, impact on health infrastructure, etc.), they operate in two different ways:

\begin{itemize}
     \item The dynamic models are meant to predict the course of the pandemic by predicting the feature data at time $t$ ($\mathbf{X}_t$) based on the data from earlier timepoints, i.e., $\mathbf{X}_0, \mathbf{X}_1, \cdots, \mathbf{X}_{t-1}$, as follows: 
    
    \begin{equation}
        \mathbf{X}_t = c + \phi_1 \times \mathbf{X}_{t-1} + \phi_2 \times \mathbf{X}_{t-2} + \cdots + \phi_p \times \mathbf{X}_{t-p} + \epsilon_t
    \end{equation}
    
    Here, $p$ is the number of past timepoints being considered, $c$ is a the constant, $\phi_i$ is the coefficient measuring the contribution of the $i^{th}$ timepoint on the prediction, and $\epsilon_t$ is the error term. This model is useful to derive a proxy feature data when the actual feature data from the current timepoint is not available.\\  
    
    \item The static models are meant to learn the relationship between the feature and decision data against observed outcomes (through a function $f_{static}$) as follows: 
    
    \begin{equation}
        \mathbf{Y} = f_{static} (\mathbf{X}, \mathbf{D}) + \epsilon
    \end{equation}
    
    \noindent These models inform each module, say the vaccine distribution module, of the outcome (in terms of infected cases, deaths, impact on healthcare infrastructure) from its past decisions based on given input factors.
\end{itemize}

\subsection{Privacy Considerations} 
\label{sec:priv} \noindent The pandemic framework operates on personal user data. We discuss the concerns over abuses of the privacy of health-related information and potential solutions to tackle them.

\subsubsection{Challenges} We first discuss some of the privacy challenges.\\ 

\noindent \textbf{Data utility vs. privacy trade-off.} Fig. \ref{fig:arc} shows that the architecture is processing data of varying privacy demand. The data from the repositories are publicly available, whereas the data reported by the mobile users are more sensitive as they pertain to personal physiological, clinical or epidemiological information. This leads to a dilemma of privacy-utility trade-off, summarized as, \textit{how much data utility can be compromised to achieve the desired data privacy guarantees and vice-versa}, and clearly it is a critical part of the privacy module (in Fig. \ref{fig:stack}) when dealing with user data. In order to achieve perfect privacy for the users, efforts have been made to model the privacy-utility trade-off as  generalizations of the information bottleneck and privacy funnel problems and come up with positive information utility under perfect privacy i.e. zero information leakage over a rate constraint \cite{Sreekumar2019Optimal}.\\ 

\noindent \textbf{Policies and regulations.} The architecture must abide by a few data-based regulations such as the General Data Protection Regulation (GDPR) or the California Consumer Privacy Act (CCPA) that are created with an objective to protect persons' right to protection of personal data from a variety of agents or organizations. These regulations deal with the following safeguards surrounding personal or sensitive data being shared using the user mobile applications.

\begin{itemize}
    \item The stakeholders (i.e., users and architecture administrator) must agree on the legal and bureaucratic facets of dissemination of personal data. This will help devise security measures to prevent either party from exploitation.
    
    \item There should be provisions for the government to induct officials who evaluate existing policies regarding the acquisition, management and utilization of user data.
    
    \item The MEC servers must host a service such that the users can view or control their own private data.
\end{itemize}

\noindent \textbf{Malicious intent and breaches.} The recommendations proposed by the architecture will potentially affect different facets of human life. It stands to reason that inaccurate predictions or recommendations may even worsen the pandemic scenario. Hence, it becomes necessary to safeguard the architecture from adversaries or threats. Depending on the privacy guarantee required, the parties could be \textit{semi-honest} or \textit{malicious}. In a semi-honest adversary model, the adversaries or parties follow their protocol definition, however, while doing so, they are allowed to learn anything from the intermediate data that is exposed to them in the defined course of computations as specified by the protocol, while in the malicious adversary model, the adversaries can deviate from the protocol definition by changing their inputs \cite{Hazay2011anote}. In either case, the cloud, MEC and mobile layers must be protected (in decreasing order of priority) from unauthorized access of individuals who can tamper with the data and computational model parameters and cause inaccurate recommendations on pandemic management.\\ 

\subsubsection{Off-the-shelf and potential solution directions} Here we propose some solutions to resolve the mentioned privacy challenges.\\ 

\noindent \textbf{Differential privacy (DP).} It is a widely-used privacy-preserving mechanism guaranteeing certain statistical bounds over loss of privacy \cite{dwork2014algorithmic}. It is a randomized query mechanism that defines a privacy loss parameter $\epsilon$. The higher the value of $\epsilon$, greater is the utility and consequent privacy loss, and vice versa. Similarly, there are  \textit{k-anonymity} techniques that operate on the idea that sensitive information about users can be obscured by combining multiple datasets with similar attributes~\cite{sweeney2002k}. The privacy layer can invoke one of these privacy-preserving techniques to control the data utility vs. privacy trade-off depending on data provenance and type.\\ 

\noindent \textbf{Privacy preserving inference.} As described in Sec. \ref{sec:arc}, the edge computing layer  provides on-demand computational services in the form of time-varying recommendations. This can be modeled as \textit{privacy-preserving inference problems}~\cite{hesamifard2018privacy, kaissis2020secure, Liu2021PPOutsourced}, involving two parties, namely mobile end users and edge devices, who respectively share and receive sensitive data. Along similar lines, Hesamifard et al. provide a framework to enable privacy-preserving machine learning as a service (MLaaS) for applying deep neural network algorithms on encrypted data~\cite{hesamifard2018privacy}. Luo et al. propose privacy-preserving clinical decision support system using Na\'ive Bayesian classifier to enable secure disease prediction and patient health status monitoring, both of which could be used in the ML modules of dynamic lockdown, pandemic prediction (see Sec. \ref{sec:fact}) and the prediction of medical requirements in critically affected regions \cite{Liu2021PPOutsourced}. Their privacy guarantees are achieved through the properties of \textit{fully homomorphic encryption} scheme and \textit{single instruction multiple data} integer circuits.\\

\noindent \textbf{Private information retrieval.} The pandemic architecture carries out information exchange in the form of queries. Queries may be \textit{inference-based} (i.e., regarding recommendations from the lockdown, mobility scheduling, prediction etc. models) or \textit{aggregation-based} (i.e., regarding gross percentage or number of vaccinated individuals in an area etc.). To achieve secure information retrieval, a class of privacy-preserving techniques, called \textit{Private Information Retrieval} (PIR), can be employed. A client in PIR can request access to records from a public repository, which may be hosted by multiple servers, without revealing information to the server(s) about the requested record \cite{chor1995Private}. One possible solution is to leverage the privacy guarantees provided by $k$-out-of-$N$ Oblivious Transfers to fetch the respective statistic of interest, thereby protecting the access pattern. Moreover, differential privacy could be applied on each category of aggregation-based query issued to protect the actual data.\\

\noindent \textbf{Hybrid privacy-preserving techniques.} These models combine several privacy-preserving mechanisms to enhance privacy and utility. For example, in order for a MEC server to compute a sum of vaccinations in a zone, all the end device nodes can encrypt their vaccination status (say, $E(1)$ for vaccinated or $E(0)$, otherwise) using some additive homomorphic cryptosystem like Paillier \cite{paillier1999publickey} or other fully homomorphic encryption system and report it to the respective MEC server. The MEC server, not knowing the private key, can then use the additive properties to obtain the encrypted aggregate value, and then add the differential privacy (DP) noise to protect the output using the homomorphic properties. Finally, the encrypted DP perturbed value is sent to the end device(s), who can then decrypt to get the actual value~\cite{uplavikar2020privacypres}.

\section{Discussions}\label{sec:disc}
\noindent We discuss the limitations of the proposed pandemic architecture and future research directions to emerge from this study.

\subsection{Limitations of the Architecture}\label{sec:lim}
\noindent The pandemic architecture suffers from several computational and data-centric challenges. \textit{First}, recall from our discussion from Sec. \ref{sec:edge_comp}, the computation at the edge relies on the big data inputs from the cloud layer as well as the mobile users. While it may be expected that the information housed in the data servers are structured, the personal and crowd-sourced data emanating from the mobile users are not. This implies that the architecture is in constant need for human intervention, i.e., an AI expert who filters the incoming data of misinformation, bias, etc. and lends structure to it, such that meaningful inferences can be derived. The domain expert may also be left to process the sizable body of scientific literature in order to ensure the method and data sources being utilized by the architecture are effective and cutting-edge~\cite{naude2020artificial}. 

\textit{Second}, the accuracy of the prediction from the AI and machine learning forecasting models hinges on steady influx of data. However, it is well-known that a significant portion of world population will not embrace digital technology for quite some time, bringing into question the pervasiveness of the proposed architecture. The adoption of the architecture is further stymied by the limitations in steady access to wireless communication technology around the world. For instance, we discuss the use of mobile applications in Sec. \ref{sec:contact} as ways to model social contagion. However, studies have shown that individuals do not always carry mobile devices or their devices are not enabled with the necessary communication technology to record contacts necessary for tracing contagion, etc.~\cite{soltani2020contact}. This suggests that the validity and reliability in prediction by the computational models is hindered due to the sporadic data influx from a small section of human population~\cite{whitelaw2020applications}. 

\textit{Third}, in contradistinction to traditional mobile computing architectures, the pandemic architecture requires decisions to be taken based on clinical parameters. Recently, the community of computational scientists are using AI for the automatic classification of COVID-19 and its physiological effects. For instance, there have been efforts to exploit computer vision to identify COVID-19 from radiological images. The radiological societies have expressed doubt that these ML models rely on features that are not necessarily related to the pathology they are classifying~\cite{lopez2021current,tseng2020computational}. Understandably, the accuracy of the architecture employing a \textit{vaccine allocation} (as discussed in ~\cite{buckner2021dynamic}) or a \textit{dynamic lockdown} recommendation (as proposed in \cite{pestieau2022optimal}) based on vision-based COVID-19 detection rests largely upon the latter's clinical robustness and generalizability. Once again, to ensure the well-being of the individual and the efficacy of the prediction model the architecture may require a bonafide medical expert in the loop to oversee the diagnostics. The downside of human participation is that it may hinder efforts to automate the architecture.

\subsection{Future Directions}\label{sec:future}
In Sec. \ref{sec:lim}, we discuss the challenges of ensuring the participation of bulk of the population and elimination of bias from the reported data. While these issues can only be addressed by long-term planning, there are a few immediate research avenues that emerge from this pandemic architecture. 

\textit{First}, a considerable amount of research effort needs to go into \textit{data sensitisation and structuring}~\cite{arora2020role}. Thankfully, the data housed in public repositories, such as the COVID Tracking Project~\cite{CTP}, of respectable organizations (refer Sec. \ref{sec:repo} for details) allow domain experts and general public to participate in the data reporting, dissemination and debugging. As described in Sec. \ref{sec:edge_comp}, the personal or crowdsourced data reported by the mobile users can be comparatively noisy and inaccurate, begging several questions on their trustworthiness and authenticity~\cite{zupanvcivc2019data}. One of the many ways to mitigate this hurdle is to assign fitness scores to mobile users based on their (1) overall reputation, (2) memory, residual energy and communication capabilities of smart devices, etc. In exchange of incentives, the system will periodically nominate the fittest individual as a \textit{cluster head} or \textit{group owner} and require him to collect the data from the peers in his neighborhood, pre-process it to eliminate potential noise, redundancy and inaccuracy, and transfer the data to the MEC nodes~\cite{vitello2018collaborative,roy2020biomcs}. Alternatively, anomaly and outlier detection mechanisms~\cite{zhang2013advancements} could be adapted at the MEC layer to bolster the accuracy of the static and dynamic prediction models in place.

\textit{Second}, as discussed in Sec. \ref{sec:priv}, there will be multiple stakeholders involved such as, government agencies, healthcare departments, police departments, third-party companies responsible for development and management tasks, users, and so on. Moreover, the data that the architecture will be collecting and using are also sensitive and confidential. Therefore, traditional hard security mechanisms that entails  authentication and authorization (access control) needs to be implemented. The objectives are to ensure only authenticated stakeholders can use the data and the services and implementation of a fine-grained authorization to enforce conflict-free separation of duties by taking into consideration spatio-temporal aspects. Henceforth, potential research directions are determination of the ``best'' authentication mechanism such that identity keys can be generated, distributed, and verified seamlessly across all platforms, and choosing a suitable access control model (viz., attribute-based access control (ABAC)~\cite{yuan2005attributed}, role-based access control (RBAC)~\cite{sandhu1998role}, organization-based access control (OrBAC)~\cite{kalam2003organization}, and so on) that enables secure accesses. Also, soft security mechanisms implemented through suitable trust/reputation models may be integrated with authentication and access control to prevent unauthorized disclosure or manipulation of information by insiders with deceptive motives.         

\textit{Third}, the most important feature of edge computing is to bring computation closer to the edge, thereby minimizing computation delay. \textit{Federated learning} -- a machine learning technique in which multiple distributed nodes can use their local data to collaboratively learn a shared prediction model \cite{XIA2021FLnECSurvey} -- may be employed to optimize this feature of edge computing. Specifically, instead of the MEC layer, the mobile devices can pool their memory and processing resources to collectively run the prediction model. While FL enables mobile devices to run applications requiring high computational power~\cite{hard2019federated}, considerable amount of research should go into ensuring efficiency, fairness and realisability of FL applications in mobile edge networks~\cite{Lim2020FLMEN,kairouz2019AdvNProbFL}.

\textit{Fourth}, yet another key aspect of the pandemic management architecture is \textit{fault-tolerance}. The mobile devices are energy constrained and prone to energy depletion, whereas the MEC and cloud servers are likely to experience hardware faults from time to time. The implications of either of these faults is data starvation leading to incorrect recommendations. Thus, research efforts need to be directed at the installation of maintenance modules that oversee the functioning at each of the three layers. There are models that improve system resilience against the intermittent connectivity of mobile devices and present preemptive measures to control the failure of cloud servers~\cite{park2011markov,bala2012fault}. The big research question then will be to integrate these models into the pandemic architecture.

\textit{Finally}, it is evident from the discussion so far that the applicability of the architecture hinges on its capability to ensure data trustworthiness and fault-tolerance. The flip side of such sophisticated mechanisms for both is the communication overhead resulting in challenges pertaining to \textit{energy efficiency}. We know that the researchers of participatory sensing and mobile computing have studied the criticality of cost-to-benefit trade-off for the mobile users at length and proposed ways for intelligent energy-efficient data reporting~\cite{tomasoni2018profiling}; additionally, the energy overheads of data centers and their consequent exorbitant demand on worldwide power consumption is also an open research area~\cite{mastelic2014cloud}. Overall, the overheads for the myriad generations of wireless communication technologies (4G, 5G, Wi-Fi, LTE, etc.) employed by the cloud, MEC and mobile layers also warrant adequate consideration in course of the realization of the architecture~\cite{chowdhury2019role}.   

\section{Conclusion}
\noindent In this paper, we proposed a pandemic management architecture that automates recommendations on vaccine distribution, dynamic lockdowns, human mobility scheduling and pandemic prediction, by leveraging the pandemic-related data collected through IoT technology. We survey the relevant computational frameworks in the field of pandemic management, online data repositories and contact-tracing mobile applications. We then delve into the wireless communication among the three layers, namely, cloud, edge and mobile layers, as well as the prediction models at the edge computational layer that run the computational models on the physiological, clinical, and epidemiological data to make time-varying recommendations. We discuss the data utility versus anonymization trade-off, privacy threats, regulations and potential solutions, before covering the limitations and future research directions to enhance the applicability of the proposed architecture.





%

\end{document}